\documentclass[usenatbib]{mnras}

\usepackage[english]{babel}
\usepackage[utf8x]{inputenc}
\usepackage[T1]{fontenc}

\usepackage{amsmath}
\usepackage{amssymb}
\usepackage{graphicx}
\usepackage{booktabs}
\usepackage{ulem}
\usepackage{gensymb}
\usepackage{url}
\usepackage[para,online,flushleft]{threeparttable}

\title[SN 1972E]{Blast From the Past: Constraining Progenitor Models of SN 1972E}
\author[Do et al.]{
Aaron Do,$^1$
Benjamin J. Shappee,$^1$
Jean-Pierre De Cuyper,$^2$
John L. Tonry,$^1$
Cynthia Hunt,$^{3, 4}$
\newauthor{
Fran\c{c}ois~Schweizer,$^3$
Mark M. Phillips,$^5$
Christopher R. Burns,$^3$
Rachael Beaton,$^{3, 6}$
}
\newauthor{
Olivier Hainaut$^7$
}
\\
$^1$Institute for Astronomy, University of Hawai`i, 2680 Woodlawn Dr., Honolulu, HI 96822, USA\\
$^2$Royal Observatory of Belgium, Ringlaan 3, B-1180 Ukkel, Belgium\\
$^3$The Observatories of the Carnegie Institution for Science, 813 Santa Barbara St., Pasadena, CA, 91101, USA\\
$^4$Giant Magellan Telescope (GMTO Corporation), 465 N Halstead St \#250, Pasadena, CA, USA 91107\\
$^5$Carnegie Observatories, Las Campanas Observatory, Casilla 601, La Serena, Chile\\
$^6$Department of Astrophysical Sciences, Princeton University, 4 Ivy Lane, Princeton, NJ, 08544, USA\\
$^7$European Southern Observatory, Karl-Schwarzschild-Strasse 2, 85748, Garching bei München, Germany}

\begin{document}
\label{firstpage}
\pagerange{\pageref{firstpage}--\pageref{lastpage}}
\maketitle

\begin{abstract}
We present a novel technique to study Type Ia supernovae by constraining surviving companions of historical extragalactic SN by combining archival photographic plates and \textit{Hubble Space Telescope} imaging.
We demonstrate this technique for Supernova 1972E, the nearest known SN Ia in 125 years.
Some models of SNe Ia describe a white dwarf with a non-degenerate companion that donates enough mass to trigger thermonuclear detonation.
Hydrodynamic simulations and stellar evolution models show that these donor stars will survive the explosion, and show increased luminosity for at least a thousand years.
Thus, late-time observations of the exact location of a supernova can constrain the presence of a surviving donor star and progenitor models.
We find the explosion site of SN 1972E by analyzing 17 digitized photographic plates taken with the European Southern Observatory 1m Schmidt and 1 plate taken with the Cerro Tololo Inter-American Observatory 1.5m telescope.
Using the \textit{Gaia} eDR3 catalog to determine Supernova 1972E's location yields: $\alpha$~=~13$^h$~39$^m$~52.708$^s$~$\pm$~0.004$^s$ and $\delta$~=~$-$31\degree~40'~8\farcs97~$\pm$~0\farcs04 (ICRS).
In 2005, \textit{HST}/ACS  imaged the host galaxy of SN 1972E with the $F435W$, $F555W$, and $F814W$ filters covering the explosion site.
The nearest detected source is offset by 3.1 times our positional precision, and is inconsistent with the colors expected of a surviving donor star.
Thus, the \textit{HST} observation rules out all Helium-star companion models and the most luminous main-sequence companion model currently in the literature.
The remaining main-sequence companion models could be tested with deeper \textit{HST} imaging.
\end{abstract}


\section{Introduction}
\label{intro}
Type Ia supernovae (SNe Ia) have become a fundamental cosmological probe for two reasons:
they release enough energy to be seen from halfway across the observable Universe \citep[e.g. ][]{rodney12}),
and their peak luminosities are monotonically correlated with their decline rate, making them standardizable candles using what is now called the Phillips relation \citep{phillips93, hamuy95, riess95}.
The distance measurements from SNe Ia can then be used to measure cosmological parameters \citep{riess98, perlmutter99}.
However, as SNe Ia-based cosmological constraints become tighter, systematic uncertainties have become the dominant source of error \citep[e.g.,][]{wood-vasey07, kessler09, guy10, conley11}.

Despite the empirical success of the Phillips relation, the exact processes involved in the explosions of SNe Ia have been difficult to identify both theoretically and observationally (for a review see \citealt{livio18}).
It is commonly accepted that a SN Ia is the runaway thermonuclear explosion of a carbon-oxygen white dwarf (WD) in a close binary system, but there is no consensus on the nature of the binary companion or the mechanism inducing the explosion.
There are two dominant models regarding the binary companion.
The first is the double degenerate (DD) scenario where the companion is also a WD.
The explosion is triggered by the merger or collision of the two WDs due to the removal of energy and angular momentum from the binary through either gravitational radiation \citep[e.g.][]{tutukov79, iben84, webbink84}, or perturbations by a third \citep[e.g.][]{thompson11, katz12, shappee13c, antognini14} or fourth \citep{pejcha13, fang18} body.
The second is the single degenerate (SD) scenario where the companion is a non-degenerate object: a main sequence (MS) star, a red giant (RG), a sub-giant, or a Helium (He) star \citep{whelan73, nomoto82}.
The explosion is triggered when one of several conditions are met:

\begin{itemize}
    \item Mass from the companion accretes onto the WD, pushing the WD towards the Chandrasekhar limit, igniting a degenerate thermonuclear runaway near the center of the WD \citep{ropke07, ma13}.
    \item He-rich matter is accreted and detonates, causing shockwaves to trigger a carbon detonation at sub-Chandrasekhar mass \citep[double detonation;][]{nomoto82, nomoto82b, nomoto84}.
    \item A rapidly spinning WD that grows to a super-Chandrasekhar mass remains stable through rotational support. If the WD loses angular momentum, it becomes unstable and explodes \citep{piersanti03, hachisu12, benvenuto15}.
\end{itemize}

Both progenitor models have been shown to lead to SNe Ia-like explosions in the literature, and both may contribute to the sum of all SNe Ia.
However, determining the fraction of SNe Ia from each channel is observationally difficult.
The DD scenario is thought to not leave behind any gravitationally bound remnant, while many versions of the SD scenario are thought to have observable consequences tied to the donor star.
Some of these versions have been ruled out as a dominant channel based on non-detections: RG companions would produce shocks as the ejecta impact the companion \citep[e.g.][]{hayden10, bianco11, bloom12, brown12} or the circumstellar material lost by the companion \citep[e.g.][]{chomiuk12, chomiuk16, horesh12, margutti12, margutti14, shappee18, cendes20}, and there would be hydrogen in the nebular phase \citep[e.g.][]{leonard07, shappee13a, maguire16, shappee18, tucker19, tucker20}.
Although non-detections abound in the literature, without positive evidence supporting the DD scenario, the best we can do is place upper limits on the prevalence of the SD scenario.
With only upper limits, many systems remain consistent with at least one version of the SD scenario (e.g. U Sco and V445 Pup, \citealt{li11}; over 20\% of the sample analyzed by \citealt{maguire13}; 3C 397, \citealt{yamaguchi15}; SN2018fhw/ASASSN-18tb, \citealt{vallely19}; SN2018cqj/ATLAS18qtd, \citealt{prieto20}).
Without knowing whether any given SN Ia is from the DD or SD scenario, it is difficult to determine what effect the progenitor system has on the observable properties of SNe Ia.
This systematic uncertainty translates directly to any SNe Ia-based cosmological constraint.

Most observations seeking to detect the signature of a SD scenario occur during the first year of evolution of the supernova, but very late time observations can also reveal the nature of the progenitor system.
In SD SN Ia explosions, both MS and He-star donors will survive the explosion, but will be shock heated and stripped of mass \citep{marietta00, pan13, shappee13}.
Despite this loss, a surviving companion star will increase in luminosity as the envelope expands.
For MS companions this overextended envelope then collapses on a Kelvin-Helmholtz time-scale, leaving the MS companions significantly more luminous (10 − $10^3$ $L_\odot$) for a long period of time ($10^3$ − $10^4$ years) but with a modestly decreased effective temperature.
When mass is lost by ablation (overheating), it is not physically possible to avoid these effects \citep{shappee13}.
Surviving He stars also increase in luminosity but remain blue \citep{pan13}.
In both of these cases, the SN ejecta is initially optically thick and will completely hide the companion for the first ∼ year, and then outshines the companion for the next $\sim 10$ years \citep{seitenzahl09, ropke12, graur16, shappee17, graur18}.
However, at even later times (e.g. > 10 years) the ejecta is both optically thin and less luminous, allowing late-time observations to directly constrain the presence of a surviving companion.

Thus far we have been restricted to search for surviving companions near the centers of supernova remnants (SNRs) to distinguish between progenitor models \citep[e.g.][]{canal01, ruiz-lapuente04, schaefer12, ruiz-lapuente18}.
Given SN Ia rates and SNR lifetimes, there are only a handful of young examples in the Galaxy and the Magellanic Clouds.
A further challenge is to precisely identify the expected location of the star.
While Type Ia SNRs are generally spherical, there are asymmetries due to the differing densities of the surrounding interstellar medium (ISM) and/or bubbles or clumps of gas and dust in the surrounding environment, which adds uncertainty to the measurement of the geometric center of the SNR \citep[e.g.][]{edwards12}.
Finally, crowding can be a significant challenge especially when looking through the Galactic disk \citep{ruiz-lapuente18}.
In this paper, we avoid these difficulties by analyzing the nearest normal SN Ia for which imaging exists, both near the time of its initial detection and 33 years after.
Beyond this paper, our technique could be applied to the handful of other close historical SNe Ia, such as SN 1895B, 1937C, 1986G, 1991T, 1991bg, and 2011fe.

\section{The Case for Supernova 1972E}
Supernova 1972E, hereafter 72E, was discovered on May 14, 1972 (UT) by \citet{kowal72} in the outer regions ($\sim 100$" from the center) of NGC 5253 as part of the Palomar Supernova Search \citep{kowal73} at a visual magnitude of $\sim 8.5$ mag.
72E was promptly classified as a Type I supernova \citep{herbig72, barbon72}.
At just 3.15 Mpc ($\mu$ = 27.49 mag; \citealt{freedman01}), 72E is still the nearest normal SN Ia since the previous SN in the same galaxy, SN 1895B \footnote{The nearest extragalactic SN Ia occurred in the Andromeda galaxy in 1885, but as a peculiar Type I, it cannot be used to study SNe Ia \citep{devaucouleurs85}}.
Furthermore, 72E became the best observed supernova with observations in the optical and infrared and was the first supernova observed with modern photoelectric detectors (e.g. \citealt{osmer72, lee72, walker72, przybylski72, cousins72, kirshner73b, ardeberg73, kirshner73a, kirshner75, vangenderen75, caldwell89}).
72E is the 4th brightest extragalactic supernova (of any type) ever observed and helped shape our understanding of supernovae physics. 

72E not only exploded in the outskirts of its host galaxy -- thus reducing photometric uncertainties due to crowding -- but it also suffered minimal host-galaxy reddening.
We fit the published light curves of 72E with the SNe Ia light curve fitter \texttt{SNooPy} \citep{burns11}, and found that 72E is consistent with having no host-galaxy reddening.
This suggests it exploded into a clean environment and, combined with its nearness, makes it an exceptional target for late-time observations.

72E provides us with a unique opportunity.
The recent SN Ia 2014J (M82), the nearest Ia since 72E, will not be a good candidate for this kind of study because it is highly obscured, located in a crowded environment, and is already showing evidence of light echoes \citep{crotts15}.
SN Ia 2011fe (M101), the brightest SN Ia since 72E, is more than twice as distant and thus the upper limits we can place on any surviving companion will be over 1.5 mag less strict.
The reason why others have not already examined 72E is because the position was not known to better than $\sim 1$" \citep{barbon08}.

High quality images of 72E are found on 17 photographic plates, taken for photometric use by Hans-Emil Schuster with the European Southern Observatory (ESO) 1m Schmidt telescope in 1973 and 1974 \citep{schuster80}, requested by Arne Ardeberg and one plate taken by James Hesser and Mario Zemelman with the Cerro Tololo Inter-American Observatory (CTIO) 1.5m telescope.
An appendix recounts the non-trivial process of locating and digitizing the astronomical plates, highlighting the importance of long-term record keeping and data archiving.

We used the upgraded Royal Observatory of Belguim (ROB) Digitiser 2 \citep{ROB12} to digitize the region around NGC 5253, archived as negative photographic densities on original ESO Schmidt (ESOS) 300mmx300mm and 0.8mm thick bended glass plates.
Each Schmidt plate was mounted on the Aerotech ABL3600 air-bearing XY-table by pressing it pneumatically between a supporting and a counter-pressure plate in order to flatten the glass for precise focusing of the emulsion.
The air bearing XY-axes had a positional stability in time in both axes of better than 35 nm.
NGC 5253 was centered in the field of view of the 12bit grey-scale CMOS BCi4-6600 camera (2208x3000 pixels of 3.5µmx3.5µm) mounted on the new purpose build two-sided 1:1 telecentric objective with no measurable (<0.1µm) optical distortion in this 7.7 mm by 10.5 mm field of view, corresponding to the same area on the plates.
With the plate-scale of 67.4 arcsec/mm of the ESO Schmidt telescope this corresponds to 236 mas/pixel and a sky coverage of 521"x708".
We used an adapted integration time to keep the sky density just below the saturation limit to maximize the dynamic grey-scale density range.
Each digitized image is composed of 32 identical camera exposures in order to minimize the illumination, rolling shutter, and shot noise.
We digitized the CTIO plate using a Gagne Porta-trace LED light panel, a Canon 5DS EOS DSLR, and a 100 mm Zeiss Makro-Planar lens.
Table \ref{tab:title} provides the details of each plate.
The digitization of ESO plate 253 is incomplete, and this plate is not included in our analysis.
In the next sections we describe how we used the plates to solve for the position of 72E in Archival \textit{Hubble Space Telescope} (\textit{HST}) images taken with the Advanced Camera for Surveys (ACS) on December 27th, 2005 (Proposal 10765; PI Zezas).

\section{Astrometry from the Digitized Plates}

Our goal in this section is to determine the equatorial coordinates of 72E using the digitized photographic plates.
In section \ref{sec:gaia} we use \textit{Gaia} eDR3 to determine the equatorial coordinates of stars near 72E.
In section \ref{sec:pixel_coords} we use a super-Gaussian function to determine the pixel coordinates of stars on the digitized plates.
In section \ref{sec:mag_eq} we correct for a systematic effect known as magnitude equation.
Lastly, in section \ref{sec:bootstrap} we combine the previous sections to calculate astrometric solutions for each plate, and transform the pixel coordinates of 72E into equatorial coordinates.

\subsection{Reference Catalog}
\label{sec:gaia}

We use \textit{Gaia} eDR3 as the reference star catalog in our astrometric solution for each photographic plate \citep{gaia16, gaia18}.
Therefore, we work in the International Celestial Reference System (ICRS).
Due to the $\sim 43$ year difference between the plate exposures and the J2015.5 epoch of \textit{Gaia} eDR3, proper motion at the $\sim 20$ mas/year level can lead to positional differences at the arcsecond level.
We use the \textit{Gaia} Archive's epoch propagation function\footnote{see \url{https://gea.esac.esa.int/archive-help/adql/epochprop/index.html} for details} to apply proper motion corrections for each star in each plate accurate to the day of the exposure.
For \textit{Gaia} eDR3 sources without proper motions this function assumes a proper motion of 0 milliarcsecond (mas) / year with an error of 0 mas / year.
These sources are not perfectly measured, but lack a proper motion and uncertainty because they failed a quality check as detailed in section 4 of \citet{lindegren18}.
For this reason, we do not use these sources in our analysis.
For the $\sim 7 000$ \textit{Gaia} eDR3 sources within a 0.4 degrees of the center of NGC 5253, the median uncertainty in RA and Dec is on the order of 0.3 mas at the J2015.5 epoch.
However, the uncertainty in proper motion is multiplied by $\sim 43$ years, and becomes the dominant source of error at $\sim 11$ mas (0.22 \textit{HST} pixels).

\subsection{Digitized Plate Centering}
\label{sec:pixel_coords}

To make use of the reference catalog, we need to determine
the pixel coordinates of each source for which we have equatorial coordinates.
We begin with initial astrometric solutions from Astrometry.net \citep{lang10}.
We transform all \textit{Gaia} eDR3 sources within 0.4 degrees of NGC 5253 using these initial solutions to identify which \textit{Gaia} eDR3 sources lie in each plate.
Many sources are saturated, necessitating a careful centering treatment.
Additionally, the noise comes from two rounds of imaging.
The first occurs during the initial exposure, when photons from the sky strike the plate to produce a negative image (dark objects on light background).
The second occurs during the digitization of the negative photographic density plate, when photons from a back-light pass through the plate.
The digitizing camera receives few photons from the opaque regions, meaning the digitized images will have the best signal-to-noise ratio (S/N) when the photon count is neither near 0, nor near the saturation limit.
We first attempted to fit each source with a 2 dimensional elliptical Gaussian, truncating any value above the saturation limit.
However, this functional form did not reproduce the smooth transition from the saturated core to the wings.
To reproduce the saturated plateau in a smooth function, we characterize each source as a super-Gaussian function with 8 free parameters.
\begin{equation}
\begin{aligned}
N(x, y) = A \exp{} & (-(a(x-x_0)^2 + 2b(x-x_0)(y-y_0)\\ 
& + c(y-y_0)^2)^P) + k \\
\end{aligned}
\end{equation}
where
\begin{eqnarray*}
a & = & \frac{\cos^2{\theta}}{2w^2_x} +  \frac{\sin^2{\theta}}{2w^2_y} \\
b & = & -\frac{\sin{2\theta}}{4w^2_x} +  \frac{\sin{2\theta}}{4w^2_y} \\
c & = & \frac{\sin^2{\theta}}{2w^2_x} +  \frac{\cos^2{\theta}}{2w^2_y}
\end{eqnarray*}

\begin{figure*}
    \centering
    \includegraphics[width=0.95\textwidth]{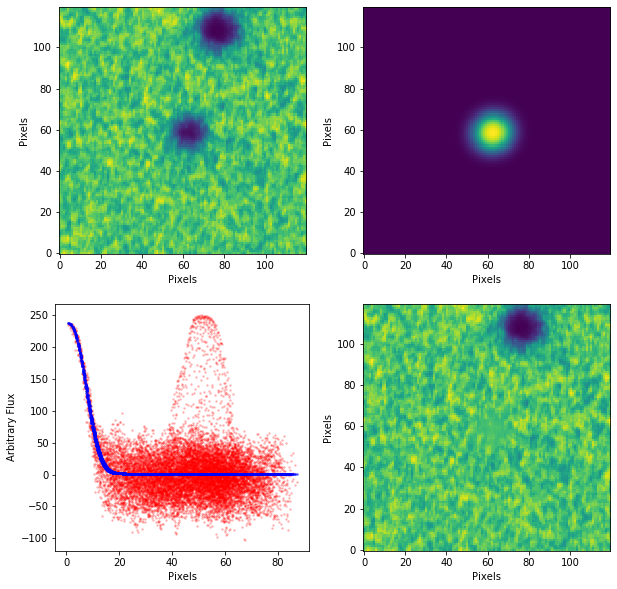}
    \caption{A randomly selected example of a successfully modeled source. Top left: 120 pixel by 120 pixel cutout of a \textit{Gaia} eDR3 source on ESO plate 328. Top right: super-Gaussian model fit. Bottom left: comparison of model (blue) and data (red) as a function of distance from the function center. The vertical axis defines 0 as the median flux with arbitrary linear scaling. Bottom right: residual upon subtraction.}
    \label{fig:centroiding}
\end{figure*}

This functional form fits for the amplitude ($A$), the center in pixel coordinates ($x_0$, $y_0$), the scale and orientation of the principal axes ($w_x$, $w_y$, $\theta$), the shape parameter ($P$), and the local sky background ($k$).
We manually evaluate each fit using diagnostic images like Figure \ref{fig:centroiding}, which shows an acceptable fit.
The residual image retains an area of low noise due to the negative imaging, but the area itself is consistent with no flux after the model is subtracted.
We reject fits where the amplitude of the star is comparable to the noise, where there are stars separated by a distance less than the recorded seeing, or where the residual image exceeds the background noise.
We do not fit sources within about an arcminute of NGC 5253's center because they would require accurate modeling and subtraction of the galaxy.
Our fits of 72E are unaffected by galaxy light.
However, there are a few issues with some of the plates.
For the CTIO plate we reject the fit for 72E because it is heavily saturated, leading to a non-zero residual.
Additionally the noise around the Northern edge of 72E in ESO plate 252 appears to be contaminating the fit, resulting in a non-zero residual.
Lastly, we reject the fits for ESO plates 489, 493, and 530 because the amplitude of the supernova is consistent with background noise in these images.

\begin{figure*}
    \centering
    \includegraphics[width=0.95\textwidth]{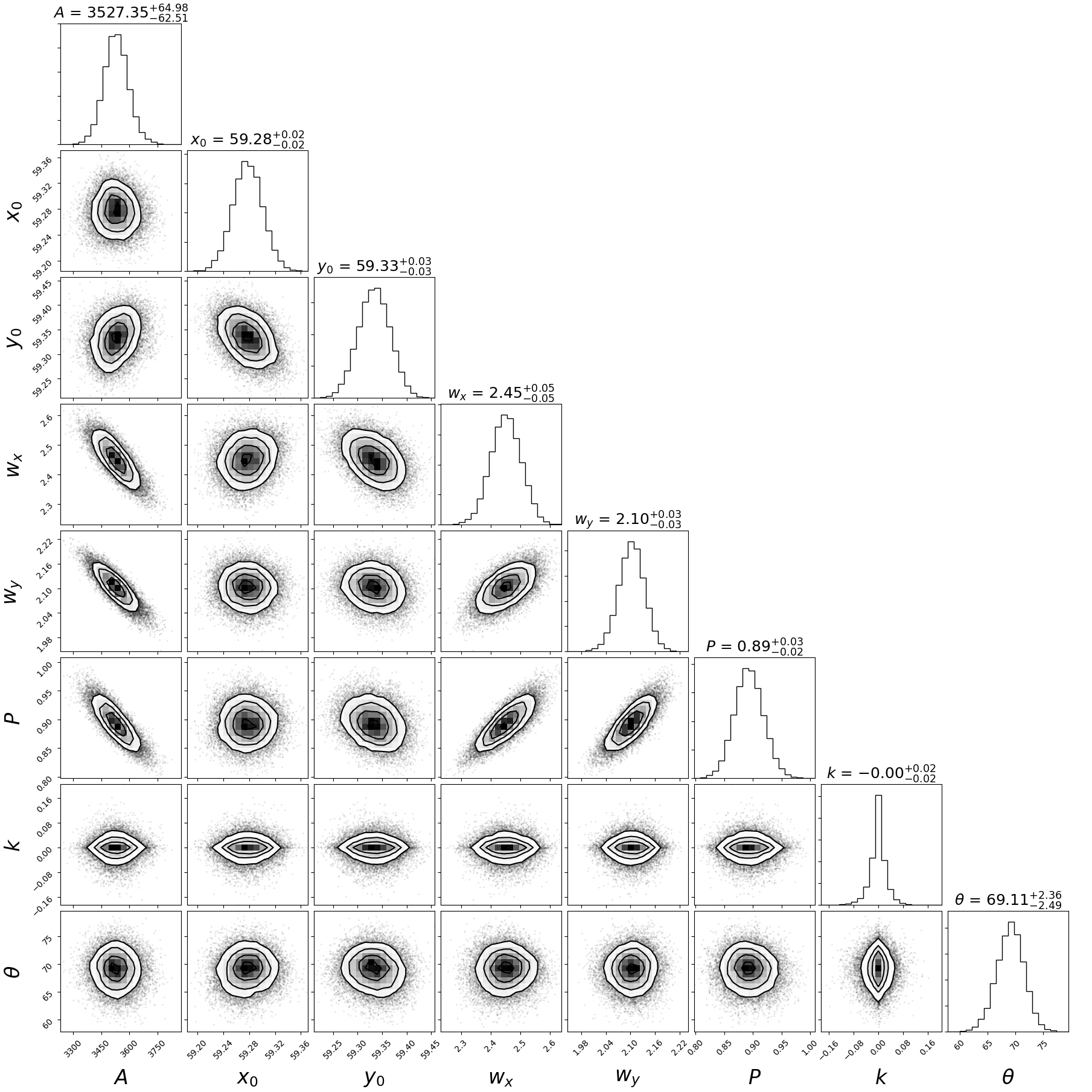}
    \caption{A corner plot demonstrating the covariances between the 8 free parameters of our model of 72E in a plate selected randomly; ESO 291. The amplitude A is relative to an arbitrary gain factor. The values of $x_0$ and $y_0$ are in pixels in a 120 pix by 120 pix cutout roughly centered on 72E. Therefore, the expected values should be near 60. The angle $\theta$ is in degrees. The autocorrelation estimator in \texttt{emcee} indicated convergence in about 6 000 steps \citep{emcee}.}
    \label{fig:corner_fit}
\end{figure*}

We quantify our uncertainty in source centering by Markov chain Monte Carlo (MCMC) sampling the posterior probability distribution of our fit using the package \texttt{emcee} \citep{emcee}.
Figure \ref{fig:corner_fit} shows the correlations between the 8 free parameters, as well as their marginalized one dimensional histograms.
$x_0$ and $y_0$, the second and third columns respectively, describe the pixel position of the source's center in a 120 pixel by 120 pixel cutout of 72E.
The figures of merit are not the values themselves, but the uncertainties of $x_0$ and $y_0$.
Our image centers are precise to $\sim 0.02$ pixels in our digitized ESO images, about 5 mas given the plate scale of 0.236 arcsec/pix, and $\sim 0.05$ pixels in our digitized CTIO images, about 8 mas given the plate scale of 0.159 arcsec/pix.

\subsection{Magnitude Equation Correction}
\label{sec:mag_eq}

In photographic plates there is a correlation between a source's intensity and its measured position on the plate.
This effect is called magnitude equation.
It means that any single astrometric solution is only accurate for sources of a magnitude similar to those used to create the solution.
It also means increasing the range of magnitudes used increases the root mean square of the residuals in the astrometric fit (RMS).
While it is possible to carefully model and correct for this effect \citep{eichhorn56, jefferys67}, we find it sufficient to restrict the stars used to those with magnitudes similar to that of 72E in the \textit{U}, \textit{B}, and \textit{V} filters of the plates.
Without correcting for magnitude equation, the RMS in each plate roughly doubles.

Although we cannot infer magnitudes directly from the 17 ESO plates because there were no sensitometer spots, and we cannot transform photographic density to intensity, we can combine photometric catalogs with estimates of 72E's magnitude at each observation epoch to determine which stars to use in each plate to construct an accurate astrometric solution.
For uniformity's sake, we use the same methodology for the CTIO plate.
The color-color transformations provided in \citet{evans18} allow us to convert \textit{Gaia} passbands into the \textit{V} passband, but there are degeneracies when trying to convert to \textit{B} magnitudes.
For that reason, we use Pan-STARRS1 \textit{g} and \textit{r} photometry to constrain the \textit{B} magnitudes of stars in the field \citep{tonry12, chambers16}.
Neither the \textit{Gaia} bandpasses, nor the Pan-STARRS1 bandpasses can be transformed into \textit{U} magnitudes, so we use the second data release of the SkyMapper Southern Survey \citep{onken19}.
SkyMapper uses a set of filters that roughly correspond to the SDSS system, except the ultraviolet band which is split into a \textit{u} and \textit{v} filter.
We find that using the SkyMapper u band as a rough proxy for the Johnson-Cousins U band allows us to infer positions for 72E consistent with the other plates, suggesting that magnitude equation is sufficiently corrected for.

We estimate 72E's magnitude at each observation epoch by fitting a univariate spline to existing photometry \citep{przybylski72, lee72, osmer72, cousins72, jarrett73, ardeberg73, ardeberg74, kirshner75, vangenderen75}.
We subtract the spline from the photometric data, and use the standard deviation of the residual magnitudes as the uncertainty for each filter.
We assume rough correspondence between UG1 glass with an -O type emulsion and the Johnson \textit{U} filter, GG385 glass with an -O type emulsion and the Johnson \textit{B} filter, and GG495 glass with a -D type emulsion and the Johnson \textit{V} filter \citep{fiorucci02, fiorucci03}.
Five of the observations were made between 415 and 763 days after maximum light, while the existing photometry only extends to about a year after maximum light.
For these five epochs we estimate 72E's magnitude using the nebular phase Phillips' relation presented in \citet{tucker20}.
The relation provides a method of approximating magnitude to within 20\% based on the peak magnitude and the decline rate.
Although the relation was calibrated with photometry from supernovae between 150 and 500 days after maximum light, the consistent linear increase in magnitude during the nebular phase of SNe Ia allows us to reasonably assume the relationship holds to the epoch of our latest plate \citep{shappee17}.

We present the estimated magnitudes of 72E in Table \ref{tab:title}.
We note that in four of the plates, 72E's magnitude is outside the magnitude range of all nearby \textit{Gaia} eDR3 sources.
Fortunately, these are CTIO 2207, and ESO plates 489, 493, and 530, plates in which we could not adequately determine the pixel coordinates of 72E.

\subsection{Determining the position of 72E and Quantifying Uncertainty}
\label{sec:bootstrap}

We compute the gnomonic projection coefficients that map the ICRS coordinates of our reference star catalogs with the pixel coordinates we measure on each plate.
We follow the Simple Imaging Polynomial (SIP) convention of \citet{shupe05}, specifying 2 $\times$ 2 matrices to account for affine transformations, and 2-dimensional complete polynomials to account for non-affine distortions.
When determining how many stars to use to calculate the astrometric solution, we considered the following factors.
\begin{itemize}
\item The effects of magnitude equation scale with the range of magnitudes used in the reference catalog, meaning we want to use a limited number of stars.

\item Higher order polynomials can model more sophisticated distortion patterns.

\item The number of coefficients in a 2-dimensional complete polynomial scales roughly with the square of the polynomial degree, increasing the number of free parameters to be fit.
\end{itemize}

To avoid overfitting, we use at least three stars for each free parameter in the distortion polynomial.
We find that a second order polynomial models the distortion field well, and with only 6 free parameters, the 18 stars we require are typically within a magnitude of 72E.
Third, fourth, and fifth order polynomials reduce the RMS, but require about 30, 45, and 63 stars respectively.
There are not enough stars with magnitudes comparable to 72E in our digitized images, meaning to avoid overfitting we would exacerbate the effects of magnitude equation.

\begin{figure*}
    \centering
    \includegraphics[width=\textwidth]{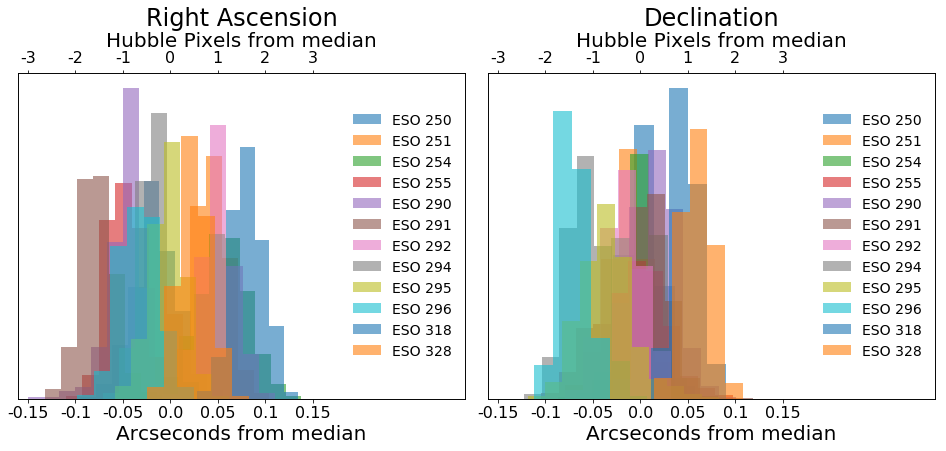}
\caption{This normalized histogram of bootstrapped coordinates shows the precision of our astrometric solutions. Each plate was resampled about 5 000 times. Left: The median right ascension is 13$^h$ 39$^m$ 52.707$^s$ (ICRS). The standard deviation is about 56 mas, or about 48 mas when multiplied by the cosine of the declination. Right: The median declination is $-$31\degree \ 40'\ 9\farcs01 (ICRS). The standard deviation is 42 mas.}
    \label{fig:hist}
\end{figure*}

To determine the ICRS coordinates of 72E, we perform a modified bootstrap resampling of the detected \textit{Gaia} eDR3 sources for each plate.
For any given plate, we give 72E a random magnitude from a normal distribution centered at the magnitude listed in Table \ref{tab:title} with a standard deviation equal to 0.12 mag in \textit{U}, 0.08 mag in \textit{B}, 0.07 mag in \textit{V}, and 0.2 mag for plates 318 and 328.
We define a bin size around each randomly generated magnitude such that at least 25 \textit{Gaia} eDR3 sources have mean \textit{G} band magnitudes within the bin.
This number is chosen somewhat arbitrarily, but attempts to provide at least 18 unique sources after selecting N sources at random with replacement, where N is the number of sources within the magnitude bin. 
We adjust each source's RA and Dec based on the uncertainties listed in the \textit{Gaia} eDR3 catalog, assuming the uncertainties are distributed normally.
We create astrometric solutions according to the measured pixel coordinates and adjusted ICRS coordinates, weighting sources according to both the inverse of their uncertainties, and how many times they were selected.
Lastly, we use these solutions to transform the detected pixel coordinates of 72E to ICRS coordinates, completing one iteration of the bootstrap resampling.
We ran about 5 000 iterations of this process for each plate.
Figure \ref{fig:hist} shows the distribution of right ascensions and declinations for the bootstrapped positions of 72E.
Each plate produces a small spread in both RA and dec.
This is a systematic uncertainty inherent to the creation of our astrometric solutions.
It describes the combined effect of the uncertainties in 72E's magnitude, the positional uncertainties in \textit{Gaia} eDR3, and the combinatoric diversity of choosing with replacement.
It is distinct from the RMS which describes the amount which any single source can deviate from the astrometric solution.

\begin{table*}
\begin{center}
\caption{Digitized Plate Derived ICRS Coordinates of SN1972E}
\label{tab:title}
\begin{threeparttable}
\tabcolsep=0.11cm
\begin{tabular}{r c c c r r c c r r}
\toprule
Plate & Date & Exposure &  Seeing & Glass & Emulsion &Equivalent Filter & Magnitude\tnote{1} & RA (ICRS) & Dec (ICRS) \\
& YY-MM-DD & minutes & & & & Johnson-Cousins & & (13:39:XX) & ($-$31:40:XX) \\
\toprule
CTIO 2207 & 72-06-17 & 30 & 1.5-2 & GG495 & ? & \textit{V}\tnote{2} & 10.21 & N/A & N/A \\ 
ESO 250 & 73-02-10 & 10 & 3-4 & GG495 & 103a-D & \textit{V} & 14.96  & $52.706$ & $09.00$ \\ 
ESO 251 & 73-02-10 & 10 & 3-4 & GG385 & 103a-O & \textit{B} & 15.12  & $52.711$ & $09.02$ \\ 
ESO 252 & 73-02-10 & 12 & 3-4 & UG1 & 103a-O & \textit{U} & 16.19  & N/A & N/A \\ 
ESO 253 & 73-02-11 & 10 & 3 & GG495 & 103a-D & \textit{V} & 14.99  & N/A & N/A \\ 
ESO 254 & 73-02-11 & 10 & 3 & GG385 & 103a-O & \textit{B} & 15.14  & $52.712$ & $09.00$ \\ 
ESO 255 & 73-02-11 & 15 & 3 & UG1 & 103a-O & \textit{U} & 16.22  & $52.704$ & $09.01$ \\  
ESO 290 & 73-03-07 & 15 & 4 & GG385 & IIa-O & \textit{B} & 15.45  & $52.704$ & $08.98$ \\ 
ESO 291 & 73-03-07 & 25 & 4 & UG1 & IIa-O & \textit{U} & 16.66  & $52.701$ & $09.00$ \\ 
ESO 292 & 73-03-07 & 15 & 3-4 & GG495 & 103a-D & \textit{V} & 15.19  & $52.711$ & $09.02$ \\ 
ESO 294 & 73-03-13 & 15 & 2 & GG495 & 103a-D & \textit{V} & 15.17  & $52.706$ & $09.06$ \\
ESO 295 & 73-03-13 & 25 & 2 & UG1 & IIa-O & \textit{U} & 16.68  & $52.707$ & $09.04$ \\ 
ESO 296 & 73-03-13 & 15 & 2 & GG385 & IIa-O & \textit{B} & 15.60  & $52.705$ & $09.07$ \\  
ESO 318 & 73-06-28 & 25 & 3 & GG385 & IIa-O & \textit{B} & 17.08  & $52.714$ & $08.95$ \\
ESO 328 & 73-07-02 & 25 & 2 & GG385 & II-O & \textit{B} & 17.14  & $52.709$ & $08.94$ \\ 
ESO 489 & 74-02-23 & 25 & 2 & GG385 & IIa-O & \textit{B} & 20.83  & N/A & N/A \\ 
ESO 493 & 74-02-25 & 40 & 4 & GG385 & 103a-O & \textit{B} & 20.87  & N/A & N/A \\
ESO 530 & 74-06-11 & 20 & 3 & GG385 & IIa-O & \textit{B} & 22.53  & N/A & N/A \\  
\bottomrule
\end{tabular}
\begin{tablenotes}
\item[1] Uncertainties are 0.12 mag in \textit{U}, 0.08 in \textit{B}, and 0.07 mag in \textit{V}. The Uncertainties in B increase to 20\%, or about 0.2 mag for plates 318, 328, 489, 493, and 530.

\item[2] GG495 glass defines the lower wavelength limit of the transmission curve, while the emulsion defines the upper wavelength limit. We assume the CTIO 2207 plate used an a-D type emulsion to produce a V filter, but we cannot be sure.
\end{tablenotes}
\end{threeparttable}
\end{center}
\end{table*}

We find our final ICRS coordinates by calculating an inverse-variance-weighted average of each plate's coordinates for 72E.
The variance comes from a number of sources added in quadrature: centering precision, \textit{Gaia} eDR3 uncertainties sampled through bootstrapping, and the RMS, all of which are listed in Table \ref{tab:error}.
The magnitude of the uncertainties varies from plate to plate, but the RMS in the astrometric solution is always the dominant source of error.
The inverse variance weighted average position is 13$^h$ 39$^m$ 52.707$^s$ $\pm$ 0.004$^s$ and $-$31\degree \ 40' \ 9\farcs01 $\pm$ 0\farcs04 (ICRS).
For each of the 14 plates in which we could determine 72E's ICRS coordinates, we present the RA and Dec determined using that plate alone in Table \ref{tab:title}.

\section{Hubble Astrometric Solution}

Our late-time observation comes from Archival \textit{HST}/ACS imaging of NGC 5253, 72E's host galaxy, taken on December 27th, 2005.
The integration times are 1759 seconds in the F435W filter, 2400s in the F555W filter, and 2360 seconds in the F814W filter.
By this date the SN was 33 years old.
The ejecta will have become optically thin after $\sim 250$ days \citep[e.g.][]{mattila05}, and the luminosity of a typical SN Ia will have declined below that of a shock-heated MS companion star after $\sim 7$ years \citep{shappee17}.

\begin{figure*}
    \centering
    \includegraphics[width=\textwidth]{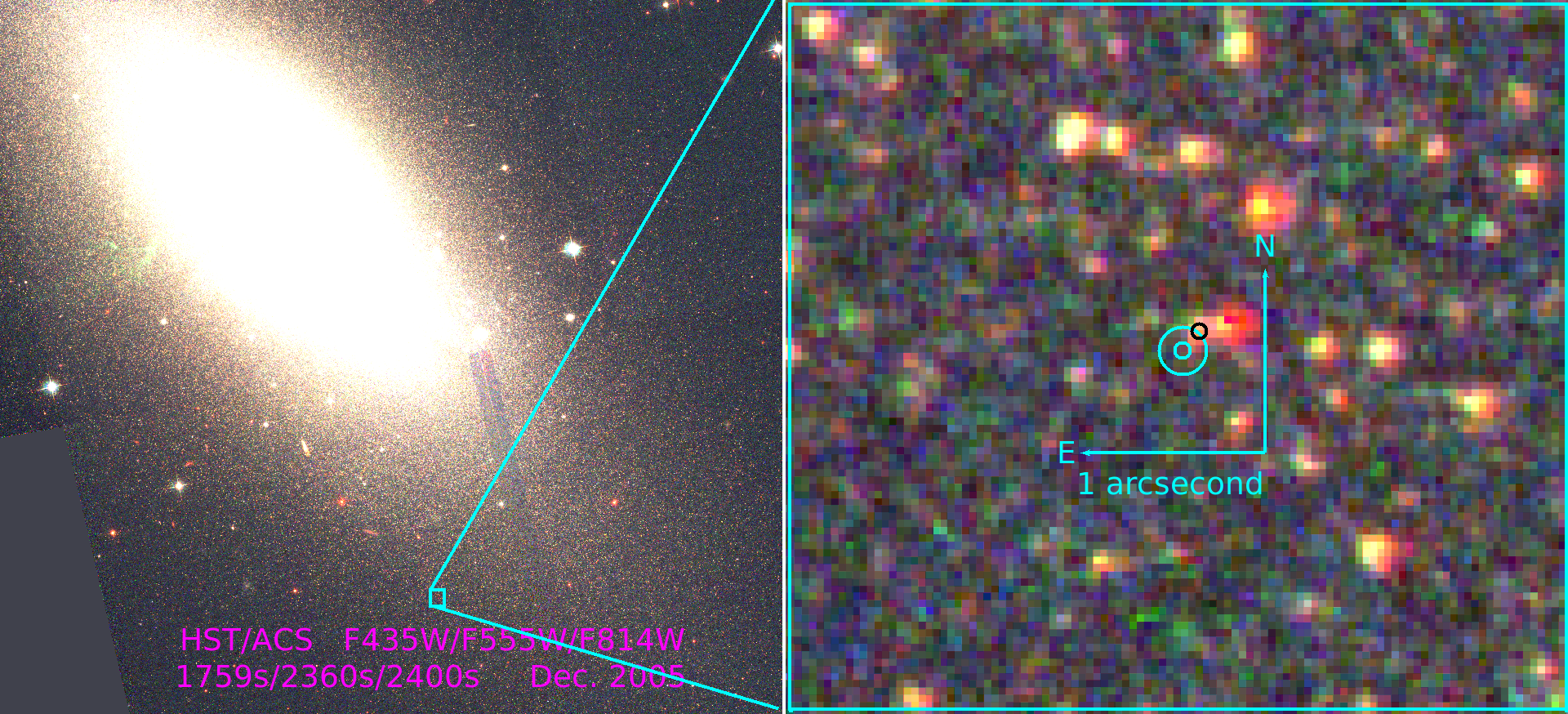}
    \caption{Archival \textit{HST}/ACS image taken at the position of SN 1972E (Proposal 10765; PI Zezas) with the location of SN 1972E and indicated by the cyan circles. Their radii are 1 and 3 times the positional uncertainty of 72E. There is no source visible to the depth of the archival image 33 years after explosion. The nearest detected source is in the black circle Northwest of 72E.}
    \label{fig:hst_no_source}
\end{figure*}

The astrometric solution from the Space Telescope Science Institute was created using the Hubble Source Catalog (HSC) version 3.1 \citep{whitmore16}.
The HSC cross-matches sources detected in \textit{HST} images with those detected in other \textit{HST} images, and those in several standard catalogs.
Notably, HSC version 3.1 was created before \textit{Gaia} eDR3 and eDR3, and does not use those catalogs.
The relative astrometric residuals between objects detected in multiple \textit{HST} ACS/WFC images have a median of 7.8 mas, and a mode of 3.4 mas.
The absolute astrometric median error is reported to be 10 mas for images near the \textit{Gaia} DR1 epoch (2015.0), increasing by 5 mas/yr for earlier and later epochs due to proper motion.
This would lead to an absolute astrometric error of about 55 mas.
We independently verify this by cross-matching the HSC sources with proper-motion subtracted \textit{Gaia} eDR3 sources within 0.2 arcseconds.
Given 183 matches between the two catalogs, the average offset would appear to be 51 mas.
However, this average includes the sources that lack proper motions and uncertainties.
We did not use these sources for our analysis.
Removing them brings the average offset from 51 mas to 38 mas, and weighting each source by its positional uncertainty brings the average offset to 32 mas.
We add this last value in quadrature to the uncertainties in RA and Dec when determining pixel coordinates.
With our calculated ICRS coordinates for 72E, the pixel coordinates of 72E are 3283.96 $\pm$ 0.95, 1714.12 $\pm$ 1.01, in the F555W filter exposure \texttt{j9k501040\_drc.chip1.fits}.
We present the pixel coordinates in the convention that assumes the center of the lower left pixel has coordinates of (0.5, 0.5) rather than (1, 1).
We show this location in the F435W/F555W/F814W false color image in Figure \ref{fig:hst_no_source}.

\begin{table*}
\begin{center}
\caption{Plate Specific Sources of Uncertainty}
\label{tab:error}
\begin{threeparttable}
\tabcolsep=0.11cm
\begin{tabular}{cccccc}
\toprule
Plate & Equivalent Filter & 72E centering & \textit{Gaia} eDR3 & Astrometric Solution & Weight \\
 & Johnson-Cousins & mas & mas & mas & Normalized to 100 \\
\toprule
ESO 250 & \textit{V} & 4.7 & 21 & 85 & 4.7\\ 
ESO 251 & \textit{B} & 3.0 & 21 & 67 & 7.2\\ 
ESO 254 & \textit{B} & 4.7 & 23 & 75 & 5.9\\ 
ESO 255 & \textit{U} & 4.7 & 26 & 67 & 6.9\\ 
ESO 290 & \textit{B} & 4.1 & 19 & 51 & 11.9\\ 
ESO 291 & \textit{U} & 5.9 & 21 & 62 & 8.3\\ 
ESO 292 & \textit{V} & 2.4 & 20 & 75 & 6.0\\ 
ESO 294 & \textit{V} & 3.5 & 18 & 59 & 9.5\\ 
ESO 295 & \textit{U} & 4.7 & 24 & 68 & 6.9\\ 
ESO 296 & \textit{B} & 4.1 & 18 & 50 & 12.5\\ 
ESO 318 & \textit{B} & 3.5 & 16 & 56 & 10.5\\ 
ESO 328 & \textit{B} & 4.7 & 16 & 58 & 9.9\\ 
\bottomrule
\end{tabular}
\end{threeparttable}
\end{center}
\end{table*}

We note that kick velocity will not affect the pixel coordinates even though companion stars of SNe Ia have been proposed as sources of hypervelocity stars \citep{wang09}.
At 3.15 Mpc, a tangential velocity of 800 km~s$^{-1}$ maintained for 33 years would cover an angular distance of only $\sim 1.3$ milliarcseconds ($0.027$ pixels).

\section{Model Constraints from Photometry}

We used \texttt{DOLPHOT} to identify sources in the F435W, F555W, and F814W images \citep{dolphin00}. and detected no source at the location of 72E.
The nearest detected source is 2.9 pixels (0.146 arcsecs) away, about 3.1 standard deviations.
It was detected at a S/N of 3.5, 7.0, and 16.1 in the F435W, F555W, and F814W filters respectively.
This source is in the Hubble Source Catalog version 3.1 under MatchID 27647143.

\begin{figure*}
    \centering
    \includegraphics[width=\textwidth]{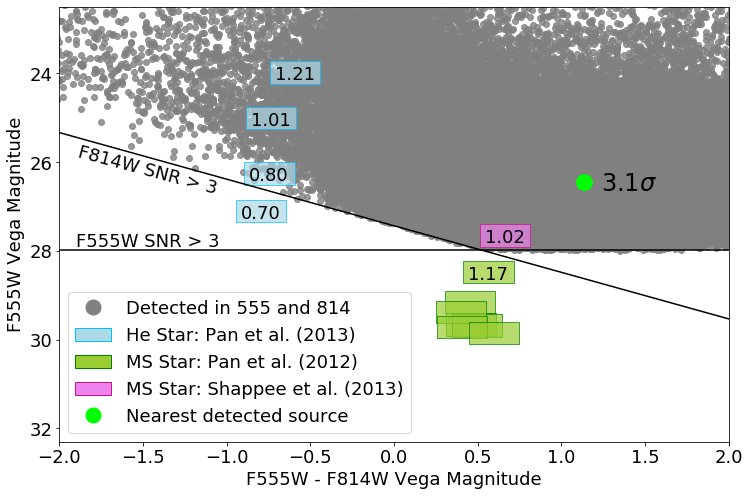}
    \caption{Color-Magnitude diagram showing what progenitor models are constrained by the Archival \textit{HST} image. The numbers in each model give the post-explosion mass of the companion star in solar masses. The Pan et al. (2012) MS stars all use a mass of 1.17 M$_\odot$. There are five models that we would have been able to detect; the Helium star models and one main sequence model. The nearest two sources are redder than any of the main sequence models.}
    \label{fig:CMD}
\end{figure*}

We derive a limiting magnitude based on the faintest stellar source detected by \texttt{DOLPHOT} (as opposed to hot pixels and elongated or extended objects).
Requiring a S/N of at least 3, the faintest sources have instrumental Vega magnitudes of 27.9 in F435W, 28.0 in F555W, and 27.4 in F814W.
Increasing the S/N requirement to 5, the faintest sources are at 27.3 in F435W, 27.4 in F555W, and 27.9 in F814W.
This rules out some progenitor models, as shown in Figure \ref{fig:CMD}.

\citet{pan13} provided evolutionary tracks for four models of He donor stars in the SD scenario.
The stars in these models have initial masses equal to 1.25, 1.35, 1.4 and 1.8 $M_\odot$, and an initial metallicity of Z = 0.02.
Thirty years after 72E, any surviving He star would be contracting from a luminous OB-like star to a hot blue-subdwarf-like (sdO-like) star.
These stars would have luminosities between 4 000 and 14 000 $L_\odot$ depending on their initial masses.
This would place all models within our limiting magnitude in the F555W filter, and all but the least massive model within our limiting magnitude in the F815W filter.
The lack of any detected source in the \textit{HST}/ACS imaging indicates that any surviving companion cannot have been a He star with a mass of 0.7 $M_\odot$ or more.

MS donor stars in the SD scenario would be overluminous, but less so than He stars.
We rule out the most luminous MS star model, which comes from \citet{shappee13}.
They explore a 1 $M_\odot$ star with four parameters: mass loss, thermal energy added, truncation radius, and the timescale of ablation and heating.
Given the set that best matched hydrodynamical simulations, a MS companion could have a luminosity of $\sim 300$ $L_\odot$ 30 years after explosion.
The apparent visual magnitude would be significantly brighter than our limiting magnitude.
Conversely, the MS donor stars modeled in \citet{pan12} reach lower luminosities ($\sim 50$ $L_\odot$).
In their models the expansion of the photosphere and increase in effective temperature may be delayed by a century or more, meaning 72E's companion would be far from its peak luminosity.

\section{Conclusion}

Our objective was to determine the presence or absence of a surviving companion for Type Ia supernova 1972E.
We have analyzed 18 digitized plates to find the RA and Dec of 72E.
We used a super-Gaussian centering kernel to find pixel coordinates and the \textit{Gaia} eDR3 catalog for equatorial coordinates.
There were six plates we did not include in our analysis: 72E was too saturated in the first plate, CTIO 2207, too faint in the last three plates, ESO plates 489, 493, and 530, and did not subtract properly in ESO plate 252.
Using the 12 remaining plates, the median values for 72E's position are $\alpha$ = 13$^h$ 39$^m$ 52.707$^s$ and $\delta$ = $-$31\degree \ 40' \ 9\farcs01 (ICRS).
We performed a bootstrap analysis to estimate our uncertainties.
The standard deviations are 48 mas in right ascension, and 42 mas in declination.
Archival \textit{HST}/ACS imaging of NGC 5253 from 2005 (Proposal 10765; PI Zezas) allows us to examine the explosion site of 72E to a depth of 27.9 mag in F435W, 28.0 mag in F555W, and 27.4 mag in F814W.
We detect no source at the given location.
The nearest source is 0.146 arcseconds away, about 3.1 standard deviations.
Furthermore, this source is more luminous than all the MS models, and redder than all the MS and He-star models described in \citet{pan12, pan13} and \citet{shappee13}.
Thus, this source is neither consistent  with the location of 72E nor the expected color from theoretical models and we rule it out as associated with 72E.

While previous studies in the literature have constrained the occurrence rate of MS companions \citep[e.g.][]{hayden10, bianco11, brown12, chomiuk16, tucker20}, the He star channel has proven more difficult to test due its blue progenitor color and high surface gravity reducing the effects of the SN ejecta interaction.
The exclusion of He-rich companions was attempted for the brightest SN Ia since 72E, SN 2011fe \citep{li11}, and the nearest since 72E, SN 2014J \citep{kelly14}.
For both supernovae, pre-explosion imaging placed upper limits on the luminosity of any companion star.
However, neither study was able to fully exclude the He-rich channel as described in \citet{liu10}.
In contrast, we are able to fully exclude this channel due to the increase in luminosity following the explosion \citep{pan13}.
\citet{tucker20} took a statistical approach to constrain SN Ia progenitor systems by looking for lines in the nebular phase spectra of 111 SNe Ia.
This method relies on radiative transfer models that have not been consistently performed for He star companions \citep{botyanszki17, botyanszki18}.

We rule out two progenitor systems.
72E was not the result of a white dwarf and a He-rich companion as modeled by \citet{pan12}.
Additionally, we rule out a MS companion as modeled by \citet{shappee13}.
However, we do not have sufficiently deep photometric data to rule out the MS models from \citet{pan12}.
The faintest of these models would have a magnitude of 29.8 in the \textit{HST}/ACS Broad V (F606W) filter which the \textit{HST} would be able to detect at a S/N of 3.5 in 10 orbits.
Looking forward, the \textit{James Webb Space Telescope} (\textit{JWST}) has 6.25 times the light gathering power of the \textit{HST} and its Near Infrared Camera (NIRCAM) has a finer scale at 32 mas pixel$^{-1}$ (compared to \textit{HST}/ACS 's 50 mas pixel$^{-1}$) \citep{perrin14}.
However, the angular resolution of NIRCAM is not significantly better than that of \textit{HST}/ACS  \citep{nelan05}.
\textit{JWST}'s key science goals involve studying high redshift, early universe phenomena, but do not necessitate increased astrometric performance.
Conversely, several of the key science cases of the \textit{Large UV/Optical/Infrared Telescope} (\textit{LUVOIR}) involve incredibly fine astrometric measurements (e.g. exoplanets causing stars to wobble, stellar proper motions due to dark matter in dwarf galaxies).
As such, the High Definition Imager (HDI) instrument has a precision astrometry mode that will be able to make sub-microarcsecond measurements \citep{luvoir}.
This far exceeds the accuracy of current astrometric catalogs, but future missions like \textit{GaiaNIR} \citep{hobbs17} are planning to improve the precision of the \textit{Gaia} eDR3 catalog by over an order of magnitude.
With the light gathering power of \textit{LUVOIR} being 40 times greater than that of the \textit{HST}, and with enough astrometric precision to mitigate the effects of crowding, we will be able to use the techniques in this paper on most SNe Ia in galaxies within $\sim$20 Mpc.
Such SNe Ia include historical supernovae like 1895B, 1937C, 1986G, 1991T, 1991bg, and 2011fe.
Applying our technique to such a sample would place firm constraints on the He star channel that has proven difficult to probe through other methods.

Understanding the progenitor systems behind SNe Ia is critical as modern surveys begin recording these transient events with increasing depth and frequency.
With systematic errors dominating the uncertainties of SNe Ia based measurements, precision cosmology depends on our ability to fully understand the process behind events like supernova 1972E.

\section*{Acknowledgments}
This work relied on the acquisition and accurate digitization of astronomical plates. For the former, we would like to thank many people: Krzysztof~Z. Stanek, Scott Gaudi, and Patrick Osmer from Ohio State University (OSU); Jos\'{e} Prieto from the Astronomy Nucleus at Universidad Diego Portales; Shrinivas Kulkarni and Jean Mueller from the California Institute for Technology (Caltech); John Johnson from Harvard University; Fernando Comer\'{o}n, and Christelle Pluciennik from the European Southern Observatory (ESO); Oscar Duhalde from the Carnegie Observatories; Dainis Dravins and Arne Ardeberg from Lund Observatory; and Steve Heathcote and Malcolm Smith from the Cerro Tololo Inter-American Observatory (CTIO).

A.D. would like to thank Michael A. Tucker, Jason T. Hinkle, Norbert Zacharias, and Chris Ashall, for reviewing this work and providing valuable feedback, Or Graur and David Rubin for very insightful conversations.
B.J.S. would like to thanks Krzysztof~Z.~Stanek, Christopher S. Kochanek, and Jennifer van Saders for useful discussion during early phases of this project.
B.J.S. is supported by NSF grants AST-1908952, AST-1920392, and AST-1911074.  

This work has made use of data from the European Space Agency (ESA) mission
{\it Gaia} (\url{https://www.cosmos.esa.int/gaia}), processed by the {\it Gaia}
Data Processing and Analysis Consortium (DPAC,
\url{https://www.cosmos.esa.int/web/gaia/dpac/consortium}). Funding for the DPAC
has been provided by national institutions, in particular the institutions
participating in the {\it Gaia} Multilateral Agreement.

The Pan-STARRS1 Surveys (PS1) and the PS1 public science archive have been made possible through contributions by the Institute for Astronomy, the University of Hawaii, the Pan-STARRS Project Office, the Max-Planck Society and its participating institutes, the Max Planck Institute for Astronomy, Heidelberg and the Max Planck Institute for Extraterrestrial Physics, Garching, The Johns Hopkins University, Durham University, the University of Edinburgh, the Queen's University Belfast, the Harvard-Smithsonian Center for Astrophysics, the Las Cumbres Observatory Global Telescope Network Incorporated, the National Central University of Taiwan, the Space Telescope Science Institute, the National Aeronautics and Space Administration under Grant No. NNX08AR22G issued through the Planetary Science Division of the NASA Science Mission Directorate, the National Science Foundation Grant No. AST-1238877, the University of Maryland, Eotvos Lorand University (ELTE), the Los Alamos National Laboratory, and the Gordon and Betty Moore Foundation.

The national facility capability for SkyMapper has been funded through ARC LIEF grant LE130100104 from the Australian Research Council, awarded to the University of Sydney, the Australian National University, Swinburne University of Technology, the University of Queensland, the University of Western Australia, the University of Melbourne, Curtin University of Technology, Monash University and the Australian Astronomical Observatory. SkyMapper is owned and operated by The Australian National University's Research School of Astronomy and Astrophysics. The survey data were processed and provided by the SkyMapper Team at ANU. The SkyMapper node of the All-Sky Virtual Observatory (ASVO) is hosted at the National Computational Infrastructure (NCI). Development and support the SkyMapper node of the ASVO has been funded in part by Astronomy Australia Limited (AAL) and the Australian Government through the Commonwealth's Education Investment Fund (EIF) and National Collaborative Research Infrastructure Strategy (NCRIS), particularly the National eResearch Collaboration Tools and Resources (NeCTAR) and the Australian National Data Service Projects (ANDS).

Based on observations made with the NASA/ESA \textit{Hubble Space Telescope}, and obtained from the Hubble Legacy Archive, which is a collaboration between the Space Telescope Science Institute (STScI/NASA), the Space Telescope European Coordinating Facility (ST-ECF/ESAC/ESA) and the Canadian Astronomy Data Centre (CADC/NRC/CSA).
\clearpage

\appendix \section{Acquiring and Digitizing the Plates}
The analysis performed in this paper was partially conceived 30 years ago in \citet{caldwell89}, where they identified a 24th magnitude object within 1 arcsecond of the location they calculated for 72E (13$^h$ 39$^m$ 52.63$^s$ $-$31\degree 40' 10\farcs1 when converted to ICRS).
The Hubble imaging revealed many more objects within that search radius.
We required a more precise estimate of 72E's coordinates, meaning we would need to create a new astrometric solution using the same plates.
So our hunt began in early 2012.

\subsection*{Palomar Discovery and 48-inch Schmidt Plates}

Charles Kowal discovered 72E on a photograph that he had obtained with the Palomar 18-inch Schmidt camera on 1972 May 14 (UT), and he confirmed it via three photographic plates that he asked one of us (F.S.) to take for him with the 48-inch Schmidt on 1972 May 16 (UT).
He announced his discovery in IAU Circular No.\ 2405 without specifying the telescopes or photographic materials used \citep{kowal72}.
The two dates he gave in that Circular (May 13 and May 15) were in local time (PDT), rather than in UT.

The discovery photograph is a circular piece of film and centered approximately on M83 (NGC~5236), a barred spiral galaxy of type SBc~II that was part of the 1972 supernova search in luminous nearby Sc galaxies \citep{kowal73}.
NGC~5253 and 72E are only $\sim$\,$1\fdg9$ away from M83 and lie well within the 8\fdg7 diameter field of view.
We know from Table~II in \citet{kowal73} that he photographed the field of M83 in 1972 March and May, with the region of the discovery film containing NGC~5253 and 72E reproduced in Plate I (Panel 5).
Kowal must have discovered 72E on this film in the afternoon or early evening of May 15, since he asked F.S. to take three short exposures of NGC~5253 with the 48-inch Schmidt right at the beginning of the night to help him confirm his discovery.
That night F.S.\ was photographing spiral galaxies for his PhD thesis project, ending a 6-night observing run initially shared with Hyron Spinrad.
If memory serves right, Kowal brought the galaxy coordinates and his plate\,+\,filter request to the 48-inch dome and then excused himself to go take another exposure of the M83 field with the 18-inch Schmidt.
Aided by Night Assistant Ranney Adams, F.S. obtained the three requested plates, choosing 72E as guide star.
Visually guiding on a supernova of $V \approx 8.7$ mag was an unforgettable experience.
Shortly thereafter, Kowal dropped by to pick up his plates and go develop them.
The three plates all bear the local date of 1972 May 15 and are: PS-17962, a 2 min exposure on Kodak 103a-O emulsion taken without filter; PS-17963, a 2 min exposure on 103a-O emulsion taken with a Schott GG13 (now called GG385) filter; and PS-17964, a 5 min exposure on 103a-D emulsion taken with a Wratten~12 filter.
Leaving the mountain the next day, F.S. never saw the developed plates.

One of these three 48-inch plates is shown in a ``Palomar Skies'' blog\footnote{\url{https://palomarskies.blogspot.com/2010/08/astrophoto-friday-supernova-1972e.html}} of 2010 Aug 6 and claimed to be the 72E discovery plate, yet it clearly is not.
Upon request by B.J.S., Scott Gaudi contacted John A.\ Johnson (at Caltech at the time) in search of an archive of Palomar plates.
Johnson indicated that Jean Mueller maintained Palomar's plate archive, verified by Shrivinas Kulkarni in conversation with B.J.S.
When B.J.S. contacted Jean Mueller, she reported that there were three 48-inch Schmidt plates, but that none of them remained in the archive.
Unfortunately, their present location and owner are unknown, but efforts to locate them will continue.

\subsection*{ESO 1-m Schmidt Plates}
Arne Ardeberg, with the help of Mart de Groot \citep{ardeberg73} immediately after the announcement of 72E on May 17, 1972 started standard UBV photometric observations at the Cassegrain focus of the 1m ESO telescope on Cerro La Silla in Chile, using a cooled EMI 6256 photomultiplier.
In early October, 1973, Arne Ardeberg convinced Hans-Emil Schuster, who was in charge of the new ESO 1m Schmidt telescope (ESOS), to take plates of NGC 5253 among those used for the testing and initial operation of the ESOS prior to the start of the ESO Southern QBS atlas program.
During the first three nights a U, B, and V filter was used and later on only a B filter in combination with different plate emulsions.
Near the end of February 1977, when Arne Ardeberg left La Silla for Lund Observatory he arranged for the 17 ESOS plates H.-E. Schuster had taken of 72E in 1973 and 1974 to be sent by ESO to Lund Observatory for him.
These plates were never used because there were no sensitometer spots exposed on the plates, hence the density to intensity transformation could not be done, and also because of problems with the standardisation using standard stars (Christiaan Sterken, 2020, Private communication).
The ESO Science Archive Facility lists twenty plates featuring NGC 5253.
Seventeen of them were taken less than two years after 72E. B.J.S. contacted Fernando Comerón about accessing the plates, but was told that they were in transit.
The ESO archive was to be moved from the ESO headquarters in Garching, Bavaria, Germany to the Royal Observatory of Belgium (ROB) in Ukkel, Brussels, Belgium.
In early 2013 B.J.S., then an astronomy graduate student at the Ohio State University, contacted Christelle Pluciennik re-inquiring as instructed.
She reached out to Jean-Pierre De Cuyper and put him in direct contact with B.J.S.

At this point a number of roadblocks emerged.
In July, J.-P. De Cuyper asked for the cabinet containing the non-atlas plates numbered from 1 to 2000 in advance of the entire shipment.
When the cabinet arrived, it did not contain the 72E plates, in contradiction with the information ESO had given beforehand.
Closer inspection of the archive revealed that the plates were on loan to '*4' and had not been returned to ESO.
The data division at ESO, one of the best observatories in archiving most completely their observations and meta data sets, had no record of persons for which non-atlas plates had been taken with the ESOS and who were marked as '*number' in the online database.
Also the maps with envelope copies of loaned Schmidt plates listed borrowers by name, not number and did not contain the 72E plates.
Thus, this lead went cold until 2016 when B.J.S. contacted J.-P. De Cuyper again hoping things had settled. J.-P. De Cuyper repeated his search, finding no new results as the logbooks that were now at ROB do not mention anything about the person for whom the 72E plates were taken.
Only for the last plates an 'SN' or one time an 'SN5253' is mentioned as object.
In desperation he redid a full search in the ESOS online database for '*4' non-survey plates, that returned 605 entries containing a '4'.
It turned out that a series of 23 plates of Carina also had been taken for '*4'. The logbook entrees for most of these plates do not mention as usual the object 'Carina' next to the coordinates, but something that resembled a name like 'Adeley' or 'Andeley'.
Only for the last 3 sets of plates taken on 1978/03/08,09,10 does the logbook show a standard entry, with the left side of the logbook mentioning an 'ARDEBERG' for the first plate of that night and next to the coordinates on the right side, the object 'Carine'.
J.-P. De Cuyper phoned Arne Ardeberg, a professor emeritus of Lund University, who replied not to know anything of SN5253 nor Carina ESO Schmidt plates.
He advised to take contact with Hans-Emil Schuster, who was in charge of the ESOS, now retired from ESO and living in his hometown Hamburg again.

Contacting Lund University, De Cuyper was directed to Dainis Dravins who immediately found both series of ESOS plates in their plate-archive still in the original ESO shipping boxes.
After waiting out the winter of 2016, Dravins shipped the two boxes to Brussels.
The digitization of the 17 plates of 72E was delayed by the upgrade of the ROB digitizer and serious illness, but was completed in 2019.
It may be natural to view historic observations as relics tracing the evolution of astronomy, but even today they can provide scientific utility through time-domain astronomy.
If there is a lesson to be learned, it is that proper data management can prevent a great deal of hardship in the future.
We encourage such efforts to be undertaken and funded.

\subsection*{CTIO Plates}
In 2016, B.J.S. contacted Patrick Osmer, who worked with CTIO data \citep{osmer72}.
Osmer could not remember what became of the plates, but directed B.J.S. to Steve Heathcote, director of CTIO.
Simultaneously, B.J.S. (at Carnegie at the time) contacted Mark Phillips who by coincidence happened to have worked with the same CTIO data in \citet{caldwell89}.
Phillips contacted Heathcote, who quickly found plate number 2207, taken by Hesser and Zemelman with the CTIO 1.5-m telescope.
B.J.S. found records of several plates taken with the Curtis Schmidt 0.6-m telescope, the CTIo 1.5 meter telescope, and the Victor M Blanco 4 meter telescope, between 1972 and 1975\footnote{Plate logs at \url{http://www.ctio.noao.edu/noao/node/189}}, but Heathcote reported large gaps in the archives surrounding their plate numbers.
He claims Osmer borrowed many of the plates in these missing regions, and Osmer suggests that he may have given them to a collaborator.
Attempts to reach out to his collaborators have not been successful.

Oscar Duhalde (Las Campanas Observatory) volunteered to create an appropriate box to use when transporting the 1.5-m plate.
After an observing run to Las Campanas Observatory B.J.S. brought the plate back to Carnegie Observatories in Pasadena to be digitized by C.H.
B.J.S then returned the CTIO plate on a subsequent observing run to Chile.

\newpage

\bibliographystyle{mnras.bst}
\bibliography{bibliography.bib}

\label{lastpage}
\end{document}